\documentclass[12pt]{article}

\usepackage{amssymb}
\usepackage{amsmath}
\usepackage{graphicx}
\usepackage{latexsym}

\begin{document}

\begin{titlepage}
\begin{center}


\hfill  OU-HET 600/2008 \\
\hfill  KEK-TH-1190 \\
\hfill  SISSA 69/2007/EP \\

\vspace{0.5cm}
{\large\bf Effective Potential of Higgs Field \\ in Warped Gauge-Higgs Unification}

\vspace{1cm}
{\bf Naoyuki Haba}$^{(a),}$
\footnote{E-mail: haba@het.phys.sci.osaka-u.ac.jp},
{\bf Shigeki Matsumoto}$^{(b),}$
\footnote{E-mail: smatsu@tuhep.phys.tohoku.ac.jp},
{\bf Nobuchika Okada}$^{(c),}$
\footnote{E-mail: okadan@post.kek.jp}, \\
and
{\bf Toshifumi Yamashita}$^{(a),(d),}$
\footnote{E-mail: yamasita@het.phys.sci.osaka-u.ac.jp}

\vspace{1cm}
{\it
$^{(a)}${Department of Physics, Osaka University,
         Toyonaka, Osaka 560-0043, Japan} \\
$^{(b)}${Tohoku University International Advanced Research and Education Organization, Institute for International Advanced Interdisciplinary Research, Sendai, Miyagi 980-8578, Japan} \\
$^{(c)}${Theory Division, KEK, 1 -1 Oho, Tsukuba, 305-0801, Japan} \\
$^{(d)}${SISSA, Via Beirut 2, I-34014 Trieste, Italy}
}

\vspace{1cm}
\abstract{
The gauge-Higgs unification is one of influential scenarios to solve the hierarchy problem in the Standard Model. Recently, the scenario on the warped background attracts many attentions due to a large possibility to construct a realistic model naturally in this framework. It is, however, well known that the effective potential for the Higgs field, which is the most important prediction of the scenario, is not easy to calculate on the warped background, because masses of Kaluza-Klein particles are not obtained analytically. In this article, we derive useful formulae for the effective potential.
The formulae allow us to calculate the Higgs mass easily, thus to
 construct a realistic model in the gauge-Higgs unification scenario on
 the warped background. Using obtained formulae, we calculate
 contributions from bulk fermions with several boundary conditions.
We also show 
 bulk fermions, which have 
 boundary conditions not allowed 
 in the orbifold picture, do not contribute
 to the effective potential. 
}


\end{center}
\end{titlepage}
\setcounter{footnote}{0}

\section{Introduction}

The Higgs sector is still a lacking piece of the Standard Model (SM). 
This sector not only governs the electroweak symmetry breaking but also
gives masses of quarks and leptons. 
Furthermore, the hierarchy problem, or to be more precise,
 the quadratic divergent correction to the Higgs mass
 strongly suggests the existence of new physics at the TeV scale.
This hierarchy problem is expected to be solved at the new physics
 scale by introducing 
 symmetries. 
For this purpose, a lot of 
 scenarios beyond the SM have been proposed so far. 
The most famous example is the supersymmety (SUSY), 
 in which quadratic divergent corrections to
 the Higgs mass term are completely cancelled. 
Another example is the little Higgs scenario, 
 where the corrections are cancelled at one-loop level
 due to an imposed (global) symmetry.  

In this article, we discuss the third possibility, so-called gauge-Higgs
 (GH) unification scenario \cite{Manton:1979kb, YH}, in which the Higgs
 mass term is controlled by higher dimensional gauge symmetry. 
In this picture,
 the Higgs field is identified as the zero mode of
 the extra dimensional component of the gauge field. 
Surprisingly, not only the quadratic divergent corrections but also the other ultraviolet (UV) divergences on the Higgs mass term completely vanish \cite{YH, Krasnikov:dt}. 
Since the imposition of the gauge invariance constrains
 the model stringently, few constructions of realistic models 
 have been performed, 
 in which the SM correctly 
 appears as the low energy effective theory of the
 model.  
This is sharp contrast to the case of SUSY or
 little Higgs scenario. 
Recently a lot of toy models of the 
 GH unification scenario have been
 considered on flat \cite{Csaki}-\cite{GGHU} and
 warped \cite{RS} backgrounds \cite{warp}-\cite{pNG}. 
Therefore, we are in the stage for the construction
 of a realistic model in the scenario. 
Once realistic model is constructed, 
it is possible to calculate signals of
 the models accurately,
 and test the idea of the GH unification at experiments such as the LHC.

There are two choices for the construction of the 
 realistic GH unification models; 
 one is  in the flat extra dimension and the other is in the 
 warped extra dimension. 
The later case seems to be attractive, because essential difficulties in
 the flat case can be resolved. 
For example, in the flat case, typical Kaluza-Klein (KK) scale, Higgs
 mass and top Yukawa coupling tend to be too small. 
These problems can be 
 naturally solved in the warped case thanks to the
 volume suppression of the gauge boson mass. 
Thus, the proper calculation of the effective potential
 for the Higgs field in the warped background is
 an important step toward the construction of realistic models.

The effective potential in the warped case, however, has been investigated less exhaustively. This situation is quite different from the flat case \cite{Lim}-\cite{Kojima:2008ky}, where even the two-loop calculation has already been performed in a certain model \cite{2loop}. The effective potential in the warped case has been evaluated in Ref.\cite{warp} for the first time. They evaluated the contribution from the gauge multiplet. A method for the calculation of the potential in a more general setup has been shown in Ref.\cite{AdS/CFT-non-orbifold}, but it is not easy to use it for the construction. 
In Ref.~\cite{PotKK}, the method in Ref.\cite{warp} was generalized to
 the case in which bulk fermions have parity-odd masses. 
Recently, more phenomenological analysis was made in Ref.~\cite{Hatanaka}. 
It was also shown in Ref.\cite{PotHolo}
 that this result can be reproduced by
 the so-called holographic approach \cite{holography}.

In this article, we review the generalization including anti-periodic fermions, and derive useful approximation formulae to calculate the effective potential easily.
We also examine contributions to the potential from fermions with non-orbifold like boundary conditions, which are often introduced in warped models \cite{AdS/CFT-non-orbifold, Hosotani:warp-non-orbifold, EWPT}. 
We show that these contributions are vanishing, although 
 such fermions seem to have non-vanishing couplings to the Higgs field. 
By applying the obtained formulae,  
 we investigate the GH condition\cite{GHC}  
 that the effective 
 theory of the GH unification model should satisfy. 

This article is organized as follows. First, in Section \ref{setup}, we clarify the setup for the derivation of formulae for the Higgs effective potential. Next, in Section \ref{Oda}, we calculate the contribution to the potential from gauge boson loop diagrams. We also explain the method how we can obtain the contribution without the knowledge of the KK mass spectrum concretely. In Section \ref{formula}, we derive formulae for contributions from bulk fermions with parity odd masses, and investigate the effective potential by using the SU(3) model as an example. 
In this section, we also examine the effects of the fermion 
 with non orbifold-like boundary conditions and 
 show that their contributions the potential are vanishing. 
Finally, we discuss the GH condition in Section \ref{Sec:GHC}. Section \ref{summary} is devoted to summary and discussions.

\section{Setup}
\label{setup}

In this article, we consider a five dimensional SU(2) model as a
simple example and derive formulae for the Higgs effective potential,
unless stated explicitly. The results can be easily applied for more
general cases in a similar way of  Ref.\cite{GeneralFormula}. In the model, the SU(2) gauge symmetry is broken down to the U(1) symmetry by the orbifold boundary conditions \cite{Kawamura}; $A_\mu^{1, 2}$ is odd, while $A_\mu^3$ is even. We calculate the effective potential of the zero-mode of $A_5^2$, setting the vacuum expectation value (VEV) of $A_5^1$ zero by using the residual U(1) symmetry.

The warped metric is taken as \cite{RS}
\begin{eqnarray}
 {\rm d} s^2
 =
 G_{M N} {\rm d} x^M {\rm d} x^N
 =
 e^{-2\sigma(y)} \eta_{\mu \nu} {\rm d} x^\mu {\rm d} x^\nu
 -
 {\rm d} y^2, 
\end{eqnarray}
where $\sigma(y) = k |y|$ at $-\pi R \leq y \leq \pi R$ and $\sigma(y) = \sigma(y + 2 \pi R)$. Four dimensional flat metric is given by $\eta_{\mu \nu} = {\rm diag}(1, -1, -1, -1)$. For clarity, we define $z(y) = e^{\sigma(y)}$, $a = 1/z(\pi R)$ and $\epsilon(y) = \sigma'(y)/k$. As a reference value, we often set the curvature $k$ and the radius $R$ to be $a_0 \equiv \exp(-\pi R k) = 10^{-15}$.

The action of gauge fields are written as 
\begin{eqnarray}
 S_g
 =
 -\frac{1}{2}
 \int \sqrt{G} {\rm d} x^4 {\rm d} y
 ~{\rm tr} \left( F_{M N} F^{M N} \right)
 =
 -\frac{1}{2}
 \int {\rm d} x^4 {\rm d} y
 ~{\rm tr} \left( F_{\mu \nu}^2 - z^{-2} F_{\mu 5}^2\right),
\end{eqnarray}
Gauge fixing term is taken to be 
\begin{eqnarray}
 S_{\rm GF}
 =
 -\int {\rm d} x^4 {\rm d} y
 ~{\rm tr} \left[ D_\mu A_\mu - D_5 \left( z^{-2} A_5\right) \right]^2,
\end{eqnarray}
where $D_{\mu, 5}$ is the covariant derivative.
 Then, we get following equations of motion,
\begin{eqnarray}
 \left( D_\mu^2 - D_5 z^{-2} D_5 \right) A_\nu = 0,
 \qquad 
 z^{-2} \left( z^2 D_\mu^2 - D_5^2 \right) \left( z^{-2} A_5 \right) = 0.
 \label{EOMA}
\end{eqnarray}
From Eq.(\ref{EOMA}), it turns out that the zero modes of $A_5$ are proportional to $z^2$. Thus, we set the VEV of $A_5^2$ as 
\begin{eqnarray}
 g \langle A_5 \rangle
 =
 g \langle A_5^2 \rangle \frac{\tau^2}{2}
 =
 A z^2 \frac{\tau^2}{2},
 \label{v}
\end{eqnarray}
where $\tau^2$ is the second Pauli matrix. It is worth notifying that this zero mode corresponds to the degree of freedom of the Wilson line phase $\theta_W$ as in the flat case,
\begin{eqnarray}
 W
 \equiv e^{i \theta_W \tau^2}
 =
 P \exp\left( {-i g \int^{\pi R}_{-\pi R} {\rm d} y ~G^{55} A_5} \right) 
 =
 \exp \left(i A \frac{1 - a^2}{2 k a^2} \tau^2 \right),
 \label{WilsonLinePase}
\end{eqnarray}
where $P$ denotes that the integration is the path ordered one. This leads to the following relation between $A$ and $\theta_W$: $A / k \equiv \widehat{A} = 2 a^2 \theta_W / (1 - a^2)$. In the following discussion, we use hatted parameters as dimensionless parameters in the unit of the curvature $k$, $e.g.$ $\widehat{A} = A/k$.

\section{Gauge Contribution}
\label{Oda}

In this section, we derive the formula for the Higgs effective potential
induced from gauge boson loop diagrams. This contribution has 
 been calculated in the Ref.\cite{warp}. 
Contributions from fermions are shown in the next section. 

\subsection{KK mass spectra}

Before going to the evaluation of the effective potential, we have to calculate the KK mass spectra of $A_\mu^1$ and $A_\mu^3$ (and also those of $A_5^1$ and $A_5^3$). Since we use the convention that only the zero mode of $A_5^2$ has a non-vanishing VEV, the classical part of the equation of motion in Eq.(\ref{EOMA}) is written as 
\begin{eqnarray}
 \left( \partial_\mu^2 - z^{-2} D_5^2 + 2 \sigma' D_5 \right) A_\nu = 0,
 \label{EOMAmuCl}
\end{eqnarray}
where $D_5 = \partial_5 - ig \langle A_5 \rangle$. With new liner combinations, $A_\mu^\pm = (A_\mu^3 \pm iA_\mu^1)/\sqrt{2}$, the covariant derivative $D_5$ can be represented diagonally as
\begin{eqnarray}
 D_5
 \begin{pmatrix}
 A_\mu^+
 \\
 A_\mu^-
 \end{pmatrix}
 =
 \begin{pmatrix}
  \partial_5 + i A z^2 & 0
  \\
  0 & \partial_5 - i A z^2
 \end{pmatrix}
 \begin{pmatrix}
  A_\mu^+
  \\
  A_\mu^-
 \end{pmatrix}.
\end{eqnarray}
This means that $\chi^\pm \equiv \exp(\mp i \int{\rm d}y A z^2) A_\mu^\pm$ satisfies the same equation of motion in Eq.(\ref{EOMAmuCl}) with the replacement of $D_5$ by $\partial_5$. Then, the equation becomes the usual equation of motion for gauge fields in the case of vanishing VEV. Using this fact, the solutions of the equation can be obtained analytically \cite{KKinRS}, 
\begin{eqnarray}
 A_\mu^\pm (x, z)
 =
 \sum_{n = 0}^\infty 
 A_{\mu n}(x)
 ~e^{\pm i \epsilon \widehat{A} z^2/2}
 ~\frac{e^\sigma}{N_n}
 \left[
  \alpha_n^\pm J_1 \left( \widehat{m}_n z \right)
  +
  \beta_n^\pm  Y_1 \left( \widehat{m}_n z \right)
 \right],
\end{eqnarray}
where we have replaced $\partial_\mu^2$ with the KK mass $m_n^2$, and assumed $\epsilon^2 = 1$. In order to obtain the canonical kinetic term for $A_{\mu n}(x)$, the normalization constant $N_n$ appears in the equation. Notice that $A_{\mu n}(x)$ is not a complex but a real field, while coefficients $\alpha_n^\pm$ and $\beta_n^\pm$ are complex. In the same manner, equations of motion for $A_5$ fields can be solved as
\begin{eqnarray}
 A_5^\pm (x, z)
 =
 \sum_{n = 0}^\infty 
 A_{5 n}(x)
 ~e^{\pm i \epsilon \widehat{A} z^2/2}
 ~\frac{e^{2\sigma}}{N_n}
 \left[
  \alpha_n^\pm J_0 \left( \widehat{m}_n z \right)
  +
  \beta_n^\pm  Y_0 \left( \widehat{m}_n z \right)
 \right],
\end{eqnarray}
where $A_5^\pm = (A_5^3 \pm iA_5^1)/\sqrt{2}$.

The KK mass spectra are determined by imposing the boundary conditions at $y = 0$ and $y = \pi R$. We have assumed that $A_\mu^1$ ($A_\mu^3$), which is proportional to ${\rm Im} A_\mu^\pm$ (${\rm Re} A_\mu^\pm$), is odd (even), 
\begin{eqnarray}
 \left.
  \frac{\rm d}{{\rm d} y}
  {\rm Re} A_\mu^\pm
 \right|_{y = 0, \pi R}
 =
 0,
 \qquad
 \left.
  {\rm Im} A_\mu^\pm
 \right|_{y = 0, \pi R}
 =0. 
\end{eqnarray}
Notice that the boundary condition for $A_\mu^-$ is the same as that for $A_\mu^+$. Thus, the conditions give four equations for two complex coefficients, $\alpha_n^\pm = (\alpha_n^3 \pm i \alpha_n^1)/\sqrt2$ and $\beta_n^\pm = (\beta_n^3 \pm i \beta_n^1)/\sqrt2$, which are summarized as 
\begin{eqnarray}
 \begin{pmatrix}
    {\cal J}_C^{\pi R}(x_n) & -{\cal J}_S^{\pi R}(x_n)
  & {\cal Y}_C^{\pi R}(x_n) & -{\cal Y}_S^{\pi R}(x_n)
  \\ 
    {\cal J}_C^{0}(x_n) & -{\cal J}_S^{0}(x_n)
  & {\cal Y}_C^{0}(x_n) & -{\cal Y}_S^{0}(x_n)
  \\ 
    s_{\widetilde{A}/2} J_1(x_n) & c_{\widetilde{A}/2} J_1(x_n) 
  & s_{\widetilde{A}/2} Y_1(x_n) & c_{\widetilde{A}/2} Y_1(x_n)
  \\ 
    s_{\widehat{A}/2} J_1(ax_n) & c_{\widehat{A}/2} J_1(ax_n) 
  & s_{\widehat{A}/2} Y_1(ax_n) & c_{\widehat{A}/2} Y_1(ax_n)
 \end{pmatrix}
 \begin{pmatrix}
  \alpha_n^3
  \\
  \alpha_n^1
  \\
  \beta_n^3
  \\
  \beta_n^1
 \end{pmatrix}
 =
 0,
 \label{BCgauge}
\end{eqnarray}
where $x_n = m_n/(ka)$, $c_x (s_x) = \cos x (\sin x)$, and $\widetilde{A} \equiv A/(ka^2) = 2\theta_W/(1-a^2)$. Functions ${\cal J}$ and ${\cal Y}$ are given by
\begin{eqnarray}
 \begin{pmatrix}
  {\cal J}_C^{0}(x)
  \\
  {\cal J}_S^{0}(x)
 \end{pmatrix}
 =
 \begin{pmatrix}
  c_{\widetilde{A}a^2/2} & -s_{\widetilde{A}a^2/2}
  \\
  s_{\widetilde{A}a^2/2} &  c_{\widetilde{A}a^2/2}
 \end{pmatrix}
 \begin{pmatrix}
  J_1(ax) + ax J'_1(ax)
  \\
  \widetilde{A}a^2 J_1(ax)
 \end{pmatrix},
\end{eqnarray}
${\cal J}_C^{\pi R}$ and ${\cal J}_S^{\pi R}$ are expressed in the same way by replacing $a$ with 1, and ${\cal Y}$ is given by these expressions with $J \rightarrow Y$. Non-zero (non-trivial) solutions exist if the determinant of the coefficient matrix, we denote it as $M(x_n; \theta_W)$, in Eq.(\ref{BCgauge}) vanishes:
\begin{eqnarray}
 N(x_n; \theta_W)
 \equiv
 {\rm det} M(x_n; \theta_W)
 =
 0.
 \label{KKMass}
\end{eqnarray}
By solving the equation for the KK mass function $N(x_n; \theta_W)$, the KK mass spectra for the $A_\mu$ field is obtained. Also the KK mass function of the $A_5$ field is obtained in the same manner, which turns out to be the same as that of $A_\mu$ due to the Higgs-like mechanism. Explicit form of the KK mass function is written as
\begin{eqnarray}
 N(x; \theta_W)
 &=&
 -\frac{ax^2}{2}
 \left[
  J_1(x)J_1(ax)Y_0(x)Y_0(ax) + J_1(x)J_0(ax)Y_0(x)Y_1(ax)
 \right.
 \nonumber \\
 &&\qquad\quad
 \left.
  + J_0(x)J_1(ax)Y_1(x)Y_0(ax) + J_0(x)J_0(ax)Y_1(x)Y_1(ax)
 \right.
 \nonumber \\
 &&\qquad\quad
 \left.
  - 2J_1(x)J_0(x)Y_0(ax)Y_1(ax) - 2J_1(ax)J_0(ax)Y_0(x)Y_1(x)
 \right]
 \nonumber \\
 &&
 + \frac2{\pi^2} \cos \left( 2\theta_W \right).
 \label{KKFnGauge}
\end{eqnarray}
An important property of $N(x;\theta_W)$ is that the function is even with respect to $x$. This fact plays an essential role in the calculation of the effective potential.

\subsection{Effective potential} 
\label{EffPot}

Once the KK mass function $N(x)$ is obtained, it is possible to calculate the one-loop effective potential
\begin{eqnarray}
 V_{\rm eff}
 =
 \sum_i \frac{1}{2}
 \int \frac{{\rm d}^4 p}{i (2 \pi)^4}
 \sum_{n = 0}^\infty \ln \left(- p^2 + m_n^2 \right),
 \label{Veff}
\end{eqnarray}
where $i$ runs over the indices of spin and gauge. We have two choices to perform the calculation. First one is that we carry out the four-momentum integral by the dimensional regularization, then replace the sum of the KK modes with the contour integral, and finally modify the path of the integrals. This method was adopted in Ref.\cite{warp} and discuss in detail, for example, in Ref.\cite{QuantumCorrection}. Second method is that we can perform the sum of the KK modes at first. For both methods, it is essential that the KK mass function is an even function. In the following, we show the calculation using the second method.

Since the KK mass function is an even function, $m_{-n} \equiv- m_n$ is also a solution in Eq.(\ref{KKMass}). Therefore, the potential in Eq.(\ref{Veff}) can be written as 
\begin{eqnarray}
 V_{\rm eff}
 =
 \sum_i \frac{1}{2} \int \frac{{\rm d}^4 p}{i (2\pi)^4}
 \frac{1}{2} \sum_{n = -\infty}^\infty
 \ln \left( - p^2 + m_n^2 \right),
\end{eqnarray}
where $n$ should run also on $-0$ for taking account of $-m_0$. The summation over $n$ is expressed by the contour integral that encircles the whole real axis counter-clockwise. After the Wick rotation, we obtain
\begin{eqnarray}
 V_{\rm eff}
 =
 \frac{3}{4} (ka)^4 \int \frac{{\rm d}^4 l}{(2\pi)^4}
 \int_C \frac{{\rm d} x}{2 \pi i}
 \ln \left( l^2 + x^2 \right) \frac{N'(x; \theta_W)}{N(x; \theta_W)},
\end{eqnarray}
where $l$ is a dimensionless Euclidean four momentum normalized by $ka$, and $N'$ denotes ${\rm d} N / {\rm d} x$. Here, we find infinite cuts on the Riemann surface in the partial integration. Since the cut starting from $x_n$ ends at $x_{-n}$, the surface term vanishes. Thus, the above equation is expressed as
\begin{eqnarray}
 V_{\rm eff}
 =
 -\frac{3}{4} (ka)^4 \int_0^\infty \frac{l^3 {\rm d} l}{8 \pi^2}
 \int_C \frac{{\rm d} x}{2 \pi i}
 \frac{(l^2 + x^2)'}{l^2+x^2}
 \ln \left[ N(x; \theta_W) \right].
\end{eqnarray}

Because of the approximation formulae of the Bessel functions at $|x| \rightarrow \infty$,
\begin{eqnarray}
 J_\nu (x)
 &\rightarrow&
 \sqrt{\frac2{\pi x}} \cos \left( x - \frac{(2\nu + 1) \pi}{4} \right),
 \\
 Y_\nu (x)
 &\rightarrow&
 \sqrt{\frac2{\pi x}} \sin \left( x - \frac{(2\nu + 1) \pi}{4} \right),
\end{eqnarray}
we find the asymptotic behavior of the KK mass function as
\begin{eqnarray}
  N(x; \theta_W)
  \rightarrow
  -\frac{2}{\pi^2}
  \left[
   \cos \left\{2 (1 - a) x \right\}
   -
   \cos \left(2 \theta_W \right)
  \right].
\end{eqnarray}
This means that, except ${\rm arg} x = 0$, the VEV dependent part of $\ln[N(x; \theta_W)]$ is exponentially suppressed at $|x| \rightarrow \infty$. Thus, it is possible to modify the contour to that encircling both upper and lower half plane clockwise, leading to the new contours encircling poles at $x = \pm i l$. After the contour integral, we obtain
\begin{eqnarray}
 V_{\rm eff}(\theta_W)
 =
 \frac{3}{2} \frac{(ka)^4}{(4\pi)^2}
 \times
 2\int_0^\infty {\rm d} x x^3
 \ln \left[ 1 + \frac{\cos(2\theta_W)}{\bar{N}_c(x)} \right].
 \label{VeffF}
\end{eqnarray}
where $\bar{N}_c(x)$ is written by the modified Bessel functions $I_\nu (x)$ and $K_\nu (x)$ as
\begin{eqnarray}
 \bar{N} (x; \theta_W)
 &\equiv&
 N(ix; \theta_W)
 \equiv
 \frac2{\pi^2}
 \left[ \bar{N}_c (x) + \cos \left( 2 \theta_W \right) \right]
 \\
 \bar{N}_c(x)
 &=&
 -\frac{2ax^2}{\pi^2}
 \left[
  I_1(x)I_1(ax)K_0(x)K_0(ax) - I_1(x)I_0(ax)K_0(x)K_1(ax)
 \right.
 \nonumber \\
 && \qquad\quad
 \left.
  - I_0(x)I_1(ax)K_1(x)K_0(ax) + I_0(x)I_0(ax)K_1(x)K_1(ax)
 \right.
 \nonumber \\
 &&\qquad\quad
 \left.
  + 2I_1(x)I_0(x)K_0(ax)K_1(ax) + 2I_1(ax)I_0(ax)K_0(x)K_1(x)
 \right].
 \nonumber
\end{eqnarray}
The result in Eq.(\ref{VeffF}) is the same as that in Ref.\cite{warp} up to the constant term.

\section{Fermion Contributions}
\label{formula}

In this section, we discuss contributions to the effective potential from bulk fermions. The strategy to calculate the contributions is essentially the same as that for the gauge contribution. We consider fermions in the fundamental representation for concreteness. Results obtained in this section can be easily extended to the case of bulk fermions in larger representations.

\subsection{Periodic fermion}
\label{Periodic Fermion}

First, we consider a periodic bulk fermion in the fundamental representation $\Psi = (\Psi_{L,R}^u, \Psi_{L,R}^d)^T$, where $\Psi_{L}^u$ and $\Psi_{R}^d$ have zero-modes. Since $\Psi_R^u$ ($\Psi_R^d$) always has the same mass as $\Psi_L^d$ ($\Psi_L^u$), we consider only $\Psi_L$ and omit the index $L$ in the following discussion. As in the case of the gauge contribution, we calculate the KK mass spectra of the fermion, and evaluate the contribution to the effective potential.

\subsubsection{KK mass spectra}

As in the case of gauge fields, the wave function of $\Psi^\pm = (\Psi^u \pm i \Psi^d)/\sqrt{2}$ with kink mass term $-i m_\Psi \bar{\Psi} \Psi$ can be obtained analytically as
\begin{eqnarray}
 \Psi^\pm (x, z)
 =
 \sum_n \Psi_n(x)
 e^{\pm\frac{i\epsilon Q \widehat{A} z^2}{2}}
 \frac{e^{\sigma/2}}{N_n}
 \left[
  \alpha_n^\pm J_\nu \left( \widehat{m}_n z \right)
  +
  \beta_n^\pm  Y_\nu \left( \widehat{m}_n z \right)
 \right],
 \label{FermionWF}
\end{eqnarray}
where $m_\Psi = c \sigma'$, $\nu = |c + 1/2|$, and $Q = 1/2$. As can be easily understood, the charge dependence of the wave function for bulk fermions in higher representations is given by $\exp[\pm i \epsilon Q \widehat{A} z^2/2] \times$ (usual solution), where $Q$ is the eigenvalue of $\tau^2$.

Boundary conditions for the bulk fermion are given by
\begin{eqnarray}
 \left.
  \left(
   \frac{\rm d}{{\rm d} y} {\rm Re} \Psi^\pm
   +
   c \sigma' {\rm Re} \Psi^\pm
  \right)
 \right|_{y = 0, \pi R}
 =
 0,
 \qquad 
 \left. {\rm Im} \Psi^\pm \right|_{y = 0, \pi R}
 =
 0,
 \label{BCPF}
\end{eqnarray}
 where the derivative might be replaced by the covariant derivative 
 so that the first condition becomes equivalent to the Dirichlet boundary 
 condition for the corresponding right-handed fermion, 
 $\left. {\rm Re} \Psi_R^\pm \right|_{y = 0, \pi R}=0$, as required by 
 the equation of motion.
Notice that this modification modifies the coefficient of the lower 
 component of the vector in the right-handed side of Eq.~(\ref{Jbar}), 
 which disappears in the determinant (\ref{Nc}) and thus does not affect 
 the effective potential.
Thus, equations for coefficients $\alpha_n^{u, d}$ and $\beta_n^{u, d}$ are obtained as
\begin{eqnarray}
 \begin{pmatrix}
    \bar{\cal J}_C^{\pi R}(x_n) & -\bar{\cal J}_S^{\pi R}(x_n)
  & \bar{\cal Y}_C^{\pi R}(x_n) & -\bar{\cal Y}_S^{\pi R}(x_n)
  \\ 
    \bar{\cal J}_C^{0}(x_n) & -\bar{\cal J}_S^{0}(x_n)
  & \bar{\cal Y}_C^{0}(x_n) & -\bar{\cal Y}_S^{0}(x_n)
  \\ 
    s_{Q \widetilde{A}/2} J_\nu(x_n) & c_{Q \widetilde{A}/2} J_\nu(x_n) 
  & s_{Q \widetilde{A}/2} Y_\nu(x_n) & c_{Q \widetilde{A}/2} Y_\nu(x_n)
  \\ 
    s_{Q \widehat{A}/2} J_\nu(ax_n) & c_{Q \widehat{A}/2} J_\nu(ax_n) 
  & s_{Q \widehat{A}/2} Y_\nu(ax_n) & c_{Q \widehat{A}/2} Y_\nu(ax_n)
 \end{pmatrix}
 \begin{pmatrix}
  \alpha_n^u
  \\
  \alpha_n^d
  \\
  \beta_n^u
  \\
  \beta_n^d
 \end{pmatrix}
 =
 0,
 \label{BC}
\end{eqnarray}
where functions $\bar{\cal J}$ and $\bar{\cal Y}$ are given by
\begin{eqnarray}
 \begin{pmatrix}
  \bar{\cal J}_C^{0}(x)
  \\
  \bar{\cal J}_S^{0}(x)
 \end{pmatrix}
 =
 \begin{pmatrix}
  c_{Q \widetilde{A}a^2/2} & -s_{Q \widetilde{A}a^2/2}
  \\
  s_{Q \widetilde{A}a^2/2} &  c_{Q \widetilde{A}a^2/2}
 \end{pmatrix}
 \begin{pmatrix}
  (1/2 + c) J_\nu(ax) + ax J'_\nu(ax)
  \\
  Q \widetilde{A}a^2 J_\nu(ax)
 \end{pmatrix},
\label{Jbar}
\end{eqnarray}
$\bar{\cal J}_C^{\pi R}$ and $\bar{\cal J}_S^{\pi R}$ are expressed in the same way by replacing $a$ with 1, and $\bar{\cal Y}$ is give by these expressions with $J \rightarrow Y$. Notice that the case with $c < 0$ can be reproduced from the case with $c > 0$ by the replacement $(\Psi_L^u, \Psi_L^d) \leftrightarrow (\Psi_R^d, \Psi_R^u)$. Thus, we consider the case with $c \geq 0$.

We define the determinant of the coefficient matrix in Eq.(\ref{BC}) as $N(x; \theta_W, c, Q)$ as in the previous section. Then, the KK mass spectrum can be found to form zeros of the determinant. After some calculations, the determinant (the KK mass function for the bulk fermion) turns out to be
\begin{eqnarray}
 N(x; \theta_W, c, Q)
 &=&
 \frac{2}{\pi^2}
  \left[ \cos \left( 2 Q \theta_W \right) + N_c(x;c)
 \right],
 \label{Nc}
 \\
 N_c(x; c)
 &=&
 1 
 +
 \frac{a x^2 \pi^2}{2}
 \left[
  J_{c + 1/2}(x) Y_{c - 1/2}(a x) -J_{c - 1/2}(a x) Y_{c + 1/2}(x)
 \right]
 \nonumber \\
 && \qquad~~~~
 \times
 \left[
  J_{c - 1/2}(x) Y_{c + 1/2}(a x) - J_{c + 1/2}(a x) Y_{c - 1/2}(x)
 \right].
 \label{NcSimple}
\end{eqnarray}
It can be seen that the $\theta_W$-dependence of $N(x; \theta_W, c, Q)$ comes only from the term $(2/\pi^2) \cos (2 Q \theta_W)$. Since the unit of $Q$ is $1/2$, $N(x; \theta_W, c, Q)$ has the sift symmetry $\theta_W \rightarrow \theta_W + 2\pi$, reflecting the phase nature of $\theta_W$. This result is consistent with that obtained in Ref.\cite{Hosotani:warp}. Below, we discuss the KK mass function $N_c(x;c)$ with some specific values of $c$, and show an approximate form of the function when $a \ll 1$.

\subsubsection*{$\bullet$ $c = 0$ case}

In the absence of the kink mass term, $c = 0$, $N_c(x; 0)$ has a simple form as
\begin{eqnarray}
 N_c(x; 0)
 =
 -\cos\left[2 (1 - a) x \right],
 \label{Nc0}
\end{eqnarray}
which leads to the KK mass, $m_n = ( n \pi \pm \theta_W/2 ) k a/(1 - a)$. Notice that the mass spectrum and therefore the effective potential are the same as those in the flat case with the replacement $1/R \leftrightarrow \pi k a/(1 - a)$.

\subsubsection*{$\bullet$ $c = 1/2$ case}

When the kink mass is given by $c = 1/2$, $N_c(x;1/2)$ is 
\begin{eqnarray}
 N_c(x; 1/2)
 &=&
 1 + \frac{a x^2 \pi^2}{2}
 \left[ J_0(x) Y_1(a x) - J_1(a x) Y_0(x) \right]
 \left[ J_1(x) Y_0(a x) - J_0(a x) Y_1(x) \right]
 \nonumber \\
 &\sim&
 1 + x \pi J_0(x) Y_1(x)
 - 2 x \left[ \gamma_E + \ln \left(\frac{ax}{2}\right) \right] J_0(x) J_1(x),
 \label{NcApp1/2}
\end{eqnarray}
where $\gamma_E$ is the Euler's constant. Notice that $N_c(x; 1/2)$ is exactly the same as the result of the gauge boson in Eq.(\ref{KKFnGauge}) when $Q = 1$. This coincidence is consistent with the fact that the gaugino field has the kink mass with $c = 1/2$ in a supersymmetric theory on the warped metric \cite{KKinRS}. The approximation in the last term is valid when $0 < x \ll 1/a$.

\subsubsection*{$\bullet$ General $c$ case}

In order to obtain a useful approximation of the KK mass function for general $c$, it is convenient to write the wave function in Eq.(\ref{FermionWF}) using $J_\nu$ and $J_{-\nu}$ as the basis of the function instead of $J_\nu$ and $Y_\nu$. After some calculations, we obtain
\begin{eqnarray}
 N_c(x; c)
 &=&
 1
 -
 \frac{\pi^2 a x^2}{2 \cos^2( c \pi )}
 \left[
  J_{1/2 + c}(x)J_{1/2 - c}(ax) + J_{-1/2 + c}(ax)J_{- 1/2 - c}(x)
 \right]
 \nonumber \\
 && \qquad\qquad~~~
 \times
 \left[
  J_{1/2 + c}(ax)J_{1/2 - c}(x) + J_{-1/2 + c}(x)J_{- 1/2 - c}(ax)
 \right].
 \label{NcSimpleJ}
\end{eqnarray}

\begin{figure}
 \begin{center}
  \scalebox{0.8}{\includegraphics*{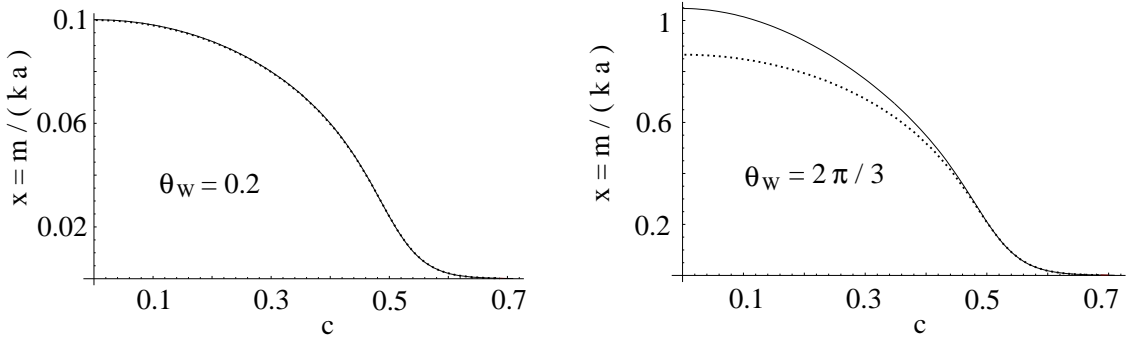}}
  \caption{\small The mass of the first KK mode with $Q=1/2$ as a function of $c$: The exact result is depicted as solid line, while the dotted one is the approximation (\ref{FirstKKApp}). We set $\theta_W = 0.2$ and $\theta_W = 2\pi/3$ in the left and right figures, respectively.}
  \label{fig:FirstKK}
 \end{center}
\end{figure}

As far as we consider a very small $a$, the contribution from $x > 1/a$ is irrelevant for our calculation. Thus, it is sufficient to know the approximate form valid for $0< x \ll 1/a$. We expand $J_\nu(a x)$ as 
\begin{eqnarray}
 J_\nu (x)
 =
 \left( \frac{x}{2} \right)^\nu
 \left[
  \frac{1}{\Gamma(\nu + 1)}
  -
  \frac{(x/2)^2}{\Gamma(\nu + 2)}
  +
  \cdots
 \right].
 \label{Japp}
\end{eqnarray}
Notice that when $-\nu + n < a$, which is realized when $|c| \sim 3/2, 5/2,\cdots$, the expansion is inappropriate. However, we are not interested in such a fine-tuned region, and  mainly interested in the case that $|c|$ is not much larger than $1/2$. Then, we find
\begin{eqnarray}
 N_c(x, c)
 =
 1
 &-&
 \frac{\pi x}{\cos(c\pi)}
 J_{- 1/2 + c}(x) J_{- 1/2 - c}(x)
 \nonumber \\
 &+&
 \frac{
  a^{1 - 2c} (2c - 1) (x/2)^{2 - 2c} \pi^2 J_{1/2 + c}(x) J_{- 1/2 + c}(x)
      }{\Gamma^2(3/2 - c) \cos^2(c \pi)},
 \label{NcApp}
\end{eqnarray}
for $c > 0$. For the $c < 0$ case, we get the same form with the replacement $c \rightarrow - c$, as it is evident from Eqs.(\ref{NcSimple}) and (\ref{NcSimpleJ}). Notice that the third term in the right-hand side of Eq.(\ref{NcApp}) is exponentially enhanced when $c - 1/2 \gg -1/\ln(a)$ due to the factor $a^{1 - 2c}$. In this case, the VEV independent part is much larger than the VEV dependent part. This leads to the quite suppressed VEV dependence on the mass spectrum, which is consistent with the fact that the coupling to the Higgs field is suppressed when $|c| > 1/2$\footnote{For $|c| > 1/2$, the left handed fermion localizes around the brane that is the opposite one where the right handed fermion localizes. Thus, the Yukawa coupling is suppressed by the small wave function overlapping.}. When $x$ is small enough, the KK mass function $N$ turns out to be
\begin{eqnarray}
 N(x; \theta_W, c, Q)
 \simeq
 \frac{\pi^2}{2}
 \left[
    \cos \left( 2 Q \theta_W \right)
  - \frac{1 - 4c^2 - 2(1 - a^{1-2c}) x^2}{1 - 4c^2}
  + {\cal O}(x^4)
 \right], 
\end{eqnarray}
which leads to the following formula for the first KK mass
\begin{eqnarray}
 m_1
 \sim
 \sqrt{\frac{(1 - 4c^2)[1 - \cos \left( 2 Q \theta_W \right)]}
            {2(1-a^{1-2c})}} k a
 \,\,\to\,\,
 \sqrt{\frac{[1 - \cos \left( 2 Q \theta_W \right)]}
            {\ln(a^{-1})}} k a, 
 \label{FirstKKApp}
\end{eqnarray}
 where the limit $c\to1/2$ is written.
This result is consistent with the approximation formula in Ref.\cite{Hosotani:warp}. In Fig.\ref{fig:FirstKK}, we show the validity of the approximation by comparing it with the exact result. As can be seen in the left figure, the approximation formula is quite consistent with the exact one. On the other hand, as shown in the right figure, the approximation becomes worse for a larger $x=m_1/(ka)$ as expected.

\begin{figure}
 \begin{center}
  \scalebox{0.9}{\includegraphics*{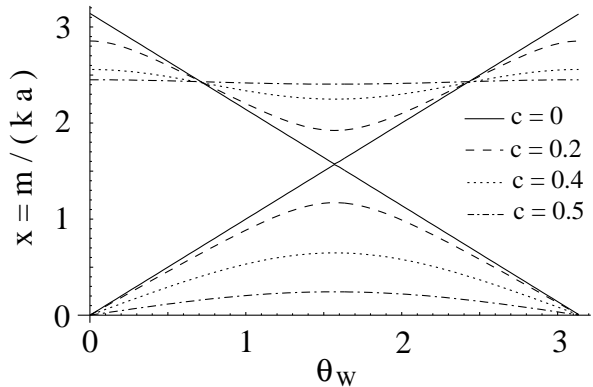}}
  \caption{\small Masses of first and the second KK modes as a function of $\theta_W$: The fermion mass is fixed at $c=0, 0.2, 0.4, 0.5$.}
  \label{fig:FirstSecondKK}
 \end{center}
\end{figure}

We also show the $\theta_W$ dependence of the first and second KK masses in Fig.\ref{fig:FirstSecondKK} with fixed $c = 0, 0.2, 0.4, 0.5$. The figure shows a similar behavior in $c \rightarrow 0$ as in the limit $k\to0$ (the flat limit) shown in Ref.\cite{Hosotani:warp}. Since gauge bosons correspond to $c = 1/2$, the W-boson mass is suppressed compared to the second KK mass.
Noting that the gauge boson, or correspondingly the left-handed fermion 
 with $c=1/2$, has a flat wave function profile along the fifth dimension, 
 we understand that their coupling is generally suppressed by the volume 
 suppression factor, $1/\sqrt{\pi R}\propto1/\sqrt{\ln(a^{-1})}$.  
On the other hand, fermions with $c=0$ which are localized around 
 the IR brane around where the Higgs fields are also localized 
 and thus free from the volume suppression.
We can see the suppression in fact appears in Eq.~(\ref{FirstKKApp}).

\subsection{Anti-periodic fermion}

Next, we consider bulk fermions obeying anti-periodic boundary conditions. These fermions are used to realize a small Wilson line phase (a small VEV of the Higgs field) \cite{HY1}, which is required to construct a realistic model consistent with the electroweak precision measurements \cite{EWPT}. Furthermore, the lightest mode of the anti-periodic fermion is stable and a good candidate for dark matter \cite{CDM}.

The equation of motion for the fermion on the warped background is the same as that of the periodic fermion, and thus the wave function is also obtained analytically using Bessel functions (\ref{FermionWF}). The difference comes only from boundary conditions; the Neumann boundary condition at $y = 0$, while the Dirichlet boundary condition at $y=\pi R$ for $\Psi^u$,
\begin{eqnarray}
 &&
 \left.
  \left(
   \frac{\rm d}{{\rm d} y} {\rm Re} \Psi^\pm
   +
   c \sigma' {\rm Re} \Psi^\pm
  \right)
 \right|_{y = 0}
 ~
 =
 0, 
 \qquad
 \left.
  {\rm Re} \Psi^\pm
 \right|_{y = \pi R}
 =
 0,
 \nonumber \\
 &&
 \left.
  \left(
   \frac{\rm d}{{\rm d} y} {\rm Im} \Psi^\pm
   +
   c \sigma' {\rm Im} \Psi^\pm
  \right)
 \right|_{y = \pi R}
 =
 0, 
 \qquad
 \left.
  {\rm Im} \Psi^\pm
 \right|_{y = 0}
 =
 0. 
 \label{BCAPF}
\end{eqnarray}
Notice that the opposite boundary condition, Dirichlet at $y = 0$ and Neumann at $y = \pi R$, gives the same result above, because it can be reproduced by the replacement, $\Psi^u \leftrightarrow \Psi^d$. Also, the case with $c < 0$ can be reproduced from the case with $c > 0$ by the replacement, $(\Psi_L^u, \Psi_L^d) \leftrightarrow (\Psi_R^d, \Psi_R^u)$. Thus, we consider the case with $c \geq 0$.

As in the case of the periodic fermion, the KK mass spectrum of the anti-periodic fermion is determined by the zeros of the determinant of the coefficient matrix,
\begin{eqnarray}
 M(x)
 =
 \begin{pmatrix}
    \bar{\cal J}_C^{\pi R}(x) & -\bar{\cal J}_S^{\pi R}(x)
  & \bar{\cal Y}_C^{\pi R}(x) & -\bar{\cal Y}_S^{\pi R}(x)
  \\ 
    -c_{Q \widehat{A}/2} J_\nu(ax) & s_{Q \widehat{A}/2} J_\nu(ax) 
  & -c_{Q \widehat{A}/2} Y_\nu(ax) & s_{Q \widehat{A}/2} Y_\nu(ax)
  \\ 
    s_{Q \widetilde{A}/2} J_\nu(x) & c_{Q \widetilde{A}/2} J_\nu(x) 
  & s_{Q \widetilde{A}/2} Y_\nu(x) & c_{Q \widetilde{A}/2} Y_\nu(x)
  \\ 
    \bar{\cal J}_S^{0}(x) & \bar{\cal J}_C^{0}(x)
  & \bar{\cal Y}_S^{0}(x) & \bar{\cal Y}_C^{0}(x)
 \end{pmatrix}.
\end{eqnarray}
Again, the determinant of the matrix, $N(x; \theta_W, c, Q) = \det M(x)$, depends on the Wilson line phase only through the term $2 \cos (2 Q \theta_W) / \pi^2$. Once we define the function $N_c(x; \theta_W)$ as in Eq.(\ref{Nc}), it is independent of $\theta_W$. The function is expressed in a simple form when $c$ is an integer or half-integer. When $c$ is not a half-integer, we can find an approximation form of it for $0 < x \ll 1/a$ by expanding the fermion wave function in terms of $J_\nu$ and $J_{-\nu}$.

Finally, we find that $N_c$ of the anti-periodic fermion is given by $- N_c$ of the periodic fermion. It means that the result of the anti-periodic fermion can be obtained from that of the periodic fermion by flipping the sign of $\cos (2 Q \theta_W)$, or equivalently by the shift, $\theta_W \rightarrow \theta_W + \pi/(2Q)$, as in the flat case \cite{HY1}. As a result, the first KK mass of the anti-periodic fermion is approximately given by
\begin{eqnarray}
 m_1
 \sim
 \sqrt{ \frac{(1 - 4c^2)[1 + \cos \left( 2 Q \theta_W \right)]}
             {2(1 - a^{1 - 2c})} } k a,
\end{eqnarray}
when $m_1 \ll k a$.

\subsection{Effective potential}

Once the KK mass function is explicitly given, the effective potential for the Higgs field is obtained using the method developed in the previous section. Then, the potential is given as
\begin{eqnarray}
 V_{\rm eff} (\theta_W; c, Q)
 &=&
 -\sum_i \frac{1}{2} \frac{(k a)^4}{(4 \pi)^2}
 v_{\rm eff} (\theta_W; c, Q)
 \nonumber \\
 v_{\rm eff} (\theta_W; c, Q)
 &=&
 2 \int_0^\infty {\rm d} x ~x^3
 \ln
 \left[
  1
  +
  \frac{\cos \left( 2 Q \theta_W \right)}
       {\bar{N}_c (x; c)}
 \right],
\end{eqnarray}
up to the cosmological constant term. Here, $i$ runs over the indices of spin, gauge and flavor. The KK mass function on the imaginary axis, $\bar{N}_c (x; c)$ is defined as
\begin{eqnarray}
 \bar{N}_c (x; c)
 &\equiv&
 N_c (ix; c)
 \nonumber \\
 &=&
 1
 -
 \frac{\pi^2 a x^2}{2 \cos^2(c \pi)}
 \left[
  I_{ 1/2 + c}(x)  I_{ 1/2 - c}(ax)
  -
  I_{-1/2 + c}(ax) I_{-1/2 - c}(x)
 \right]
 \nonumber \\
 && \qquad\qquad~~~
 \times
 \left[
  I_{ 1/2 + c}(ax) I_{ 1/2 - c}(x)
  -
  I_{-1/2 + c}(x)  I_{-1/2 - c}(ax)
 \right].
 \nonumber \\
 &\sim&
 1
 -
 \frac{\pi x}{\cos(c \pi)} I_{-1/2 + c}(x) I_{-1/2 - c}(x)
 \nonumber \\
 && ~~
 -
 \frac{a^{1 - 2c}(2c - 1) (x/2)^{2 - 2c} \pi^2 I_{1/2 + c}(x) I_{-1/2 + c}(x)}
      {\Gamma^2(3/2 - c) \cos^2(c \pi)}.
 \label{NcbarApp}
\end{eqnarray}
In the derivation of the approximation formula, we have used the relation, $J_\nu (i x) = i^\nu I_\nu (x)$ and the expansion in Eq.(\ref{Japp}). In the limit $c \rightarrow 1/2$, above approximate formula gives
\begin{eqnarray}
 \bar{N}_c (x; 1/2)
 \sim
 1
 -
 2 x I_0(x) K_1(x)
 +
 2 x
 \left[
  \gamma_E
  +
  \ln \left( \frac{ax}{2} \right)
 \right]
 I_0(x) I_1(x).
\end{eqnarray}

Here, we calculate the Higgs mass from the effective potential obtained. In order to do that, we have to know the relation between $A_5$ and the four dimensional Higgs field $h$. We set $A_5^2 = N^{-1} h z^2$, where $N$ is the normalization factor to have the canonical kinetic term after the integration of the 5th dimension. From the decomposition, we find
\begin{eqnarray}
 {\cal L}_H
 =
 -\int_{- \pi R}^{\pi R} \sqrt{G} {\rm d} y
 \frac{1}{2} {\rm tr}
 \left( F_{MN} F_{PQ} \right) G^{MN} G^{PQ}
 \ni
 \int_{- \pi R}^{\pi R} {\rm d} y z^2 N^{-2}
 \frac{1}{2} \left( \partial_\mu h \right)^2,
\end{eqnarray}
therefore, $N = [(1 - a^2)/(k a^2)]^{1/2}$. With the definition of $A$ in Eq.(\ref{v}) and the effective four dimensional gauge coupling $g_4^2 = g^2/(2 \pi R)$, we see $\partial_h^2 = - g_4^2 (1 - a^2) \ln (a) / (2 k^2 a^2) \partial_{\theta_W}^2$. Therefore, the Higgs mass is given by
\begin{eqnarray}
 m_h^2
 =
 \frac{ {\rm d}^2 V_{\rm eff} }{ {\rm d} h^2}
 =
 - g_4^2 \ln(a) \frac{(k a)^2}{64 \pi^2}
 \sum_i d_i \sum_j
 \frac{ {\rm d}^2 v_{\rm eff} }{ {\rm d} \theta_W^2 },
 \label{HiggsMass}
\end{eqnarray}
where, $i (j)$ runs over the flavor (gauge) index of each multiplet, and $d_i$ is the number of the spin degree of freedom; for example, $d_i = 3$ for the gauge multiplet and $d_i = - 4$ for Dirac fermions. Notice that the W boson mass is estimated from Eq.(\ref{FirstKKApp}) in the limit $c \rightarrow 1/2$ as $m_W^2 = -(1 - \cos\theta_W) (ka)^2 / \ln a$, which is sufficiently smaller than $k a$, thus the approximation is reliable. Then, we obtain
\begin{eqnarray}
 m_h^2
 =
 \frac{g_4^2 (\ln a)^2}{64 \pi^2 (1 - \cos\theta_W)} m_W^2
 \sum_i d_i \sum_j
 \frac{ {\rm d}^2 v_{\rm eff} (\theta_W; c_i, Q_a)}{ {\rm d} \theta_W^2 }.
 \label{HiggsMassPhys}
\end{eqnarray}

\begin{figure}
 \begin{center}
  \scalebox{0.75}{\includegraphics*{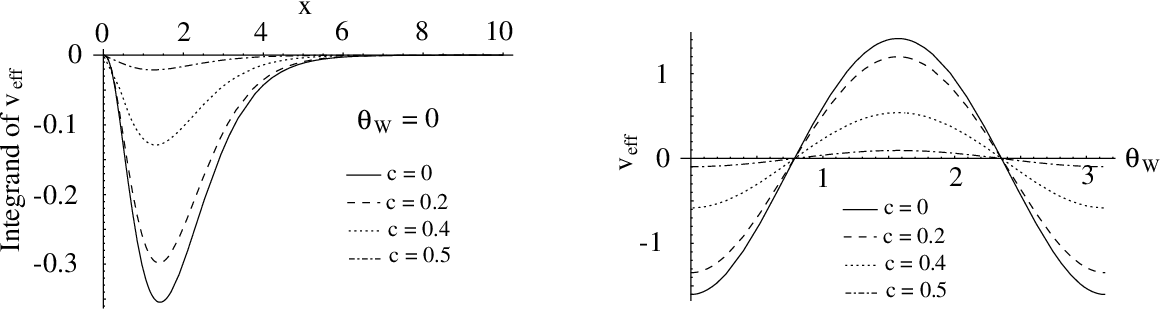}}
  \caption{\small The $c$ dependence of the integrand of $v_{\rm eff}$ (left figure) and $v_{\rm eff}$ (right figure): In the left figure, we set $\theta_W = 0$.}
  \label{fig:LogF}
 \end{center}
\end{figure}

The $c$ dependence of the integrand and that of $v_{\rm eff}$ are shown
 in Fig.\ref{fig:LogF}. We find that the effective potential is
 essentially determined by the contribution from the region $x \sim1$,
 and other regions, $x \ll 1$ and $x \gg 1$, are negligible. Also, we
 can see that the contribution to 
 the effective potential is smaller for the larger $c$ in the right figure, which is expected from the fact that the Yukawa coupling becomes smaller.

\subsection{A model example}
\label{Sec:ModelExample}

\begin{figure}
 \begin{center}
  \scalebox{0.9}{\includegraphics*{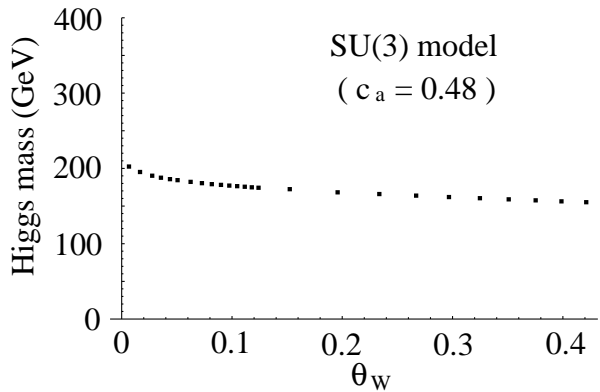}}
  \caption{\small The relation between the Higgs mass and the value of $\theta_W$ at the vacuum in the SU(3) model: Here, we set $m_W = 80$ GeV, $a = a_0 = 10^{-15}$, and $g_4 = 0.6516$.}
  \label{fig:HiggsMass}
 \end{center}
\end{figure}

In this subsection, we examine the Higgs mass in a concrete model using
formulae derived in this and previous sections. 
For example, we consider the SU(3) model, where the SU(3) symmetry is broken down to the SU(2) $\times$ U(1) by the orbifold breaking. We assume that the components of $A_5$ corresponding to the SU(3)/SU(2) $\times$ U(1) symmetry have zero modes. We introduce a pair of the fundamental fermions with the anti-periodic boundary condition and an adjoint fermion with the periodic one in this model. Their parity odd masses are set to be $c_f$ and $c_a$, respectively. Then, we find that the effective potential of the Higgs field turns out to be \cite{GeneralFormula}
\begin{eqnarray}
 V_{\rm eff} (\theta_W)
 &=&
 \frac{(ka)^4}{2(4\pi)^2}
 \left[
  3
  \left\{
    v_{\rm eff} \left(\theta_W; \frac{1}{2}, 1\right)
   +
   2v_{\rm eff} \left(\theta_W; \frac{1}{2}, \frac{1}{2}\right)
  \right\}
 \right.
 \\
 &&
 \left.
  -8v_{\rm eff} \left(\theta_W + \pi; c_f, \frac{1}{2}\right)
  -4
  \left\{
    v_{\rm eff} \left(\theta_W; c_a, 1\right)
    +
   2v_{\rm eff} \left(\theta_W; c_a, \frac{1}{2}\right)
  \right\}
 \right],
 \nonumber
\end{eqnarray}
Here we fix the mass of the periodic fermion as $c_a = 0.48$, because the critical value of $c_f$ to realize the SU(2) $\times$ U(1) symmetry breaking is $c_f^0 \sim 0.447$. In Fig.\ref{fig:HiggsMass}, the relation between the Higgs mass defined in Eq.(\ref{HiggsMassPhys}) and the value of $\theta_W$ at the vacuum is shown by tuning $c_f$ appropriately. We see that the Higgs mass is certainly large compared to $m_W$, which is in sharp contrast to the flat case.

\subsection{Non orbifold-like fermions}
\label{Sec:NonOrbifold}

In the SU(2) model, there are four orbifold-like fermions, which obey the conditions,
\begin{eqnarray}
 \begin{pmatrix}
  N, N \\
  D, D
 \end{pmatrix},
 \quad
 \begin{pmatrix}
  D, D \\
  N, N
 \end{pmatrix},
 \quad
 \begin{pmatrix}
  N, D \\
  D, N
 \end{pmatrix},
 \quad
 \begin{pmatrix}
  D, N \\
  N, D
 \end{pmatrix},
\label{OrbifoldBC}
\end{eqnarray}
where $N$ ($D$) indicates the Neumann (Dirichlet) condition, the upper (lower) signs show conditions for $\Psi^u$ ($\Psi^d$), and the left side (right side) signs show conditions at the boundary on $y = 0$ ($y = \pi R$). Notice that these conditions are for $\Psi_L$, and those for $\Psi_R$ have opposite conditions. Therefore, the first (third) boundary condition becomes the second (fourth) one, if we exchange $\Psi^u$ and $\Psi^d$. As a result, the second (fourth) condition gives the same contribution as the first (third) one to the effective potential. As can be seen in Eqs.(\ref{BCPF}) and (\ref{BCAPF}), the first and third conditions are nothing but those for the periodic and anti-periodic fermions, respectively.

Here, we investigate other possibilities,
 that is, 
 non orbifold-like boundary conditions. 
Such conditions are not consistent with the orbifold picture, but still allowed if we regard the extra dimension as an interval \cite{Higgsless}. Once we assign the opposite sign for $\Psi_R$ against $\Psi_L$, there are $2^{2^2} = 16$ possibilities, which comes from two choices of Neumann or Dirichlet conditions at two boundaries $y = 0$ and $\pi R$ for two components $\Psi^u$ and $\Psi^d$. Among them, the above four are orbifold-like, and other four conditions have the same forms as those of four conditions in remaining eight possibilities by exchanging $\Psi^u$ and $\Psi^d$. Therefore, there are eight possibilities to examine,
\begin{eqnarray}
\small{
 \begin{pmatrix}
  N, N \\
  N, N
 \end{pmatrix},
 \begin{pmatrix}
  N, N \\
  N, D
 \end{pmatrix},
 \begin{pmatrix}
  N, N \\
  D, N
 \end{pmatrix},
 \begin{pmatrix}
  N, D \\
  N, D
 \end{pmatrix},
 \begin{pmatrix}
  D, N \\
  D, N
 \end{pmatrix},
 \begin{pmatrix}
  D, D \\
  D, N
 \end{pmatrix},
 \begin{pmatrix}
  D, D \\
  N, D
 \end{pmatrix},
 \begin{pmatrix}
  D, D \\
  D, D
 \end{pmatrix}
}.
\label{NonOrbifoldBC}
\end{eqnarray}

These non orbifold-like boundary conditions can be effectively realized 
 from the orbifold-like ones by introducing boundary localized 
 (chiral and SU(2)-breaking) 
 fermions with infinitely large mixing masses to
 the bulk fermions\cite{Haba:2008hq}. 
For instance, let us 
 start from the first option in Eq.~(\ref{OrbifoldBC}), which 
 implies the right-handed partner has the second option. 
Then we introduce 
 boundary fields ${\psi_R^u}^0$ and ${\psi_L^d}^0$ on $y=0$ and 
 ${\psi_R^u}^{\pi R}$ and ${\psi_L^d}^{\pi R}$ on $y=\pi R$ to compose localized 
 mixing mass terms:
\begin{eqnarray}
 & & \left(\sqrt{2m_L^0}\bar{\psi_R^u}^0\Psi_L^u
          +\sqrt{2m_R^0}\bar{\psi_L^d}^0\Psi_R^d
     \right)\delta(y)
\nonumber\\
 &+& \left(\sqrt{2m_L^{\pi R}}\bar{\psi_R^u}^{\pi R}\Psi_L^u
          +\sqrt{2m_R^{\pi R}}\bar{\psi_L^d}^{\pi R}\Psi_R^d
     \right)\delta(y-\pi R).
\end{eqnarray}
When we take a limit $m_R^{0,\pi R}\to\infty$ (without introducing 
 ${\psi_R^u}^{0,\pi R}$), the wave function of $\Psi_R^d$ vanishes 
 at both boundaries to avoid 
 the large mass terms. 
This means the boundary conditions of $\Psi_R^d$ become Dirichlet at both 
 the boundaries, in which the boundary conditions of $\Psi_L^d$
 change to 
 the Neumann due to the equation of motion.
Thus, the first boundary condition 
 in Eq.~(\ref{NonOrbifoldBC}) is realized.
In the same way,  other boundary conditions in Eq.~(\ref{NonOrbifoldBC}) 
 can be realized from orbifold-like boundary conditions. 

To be more concrete, the above localized mass terms modify the boundary 
 conditions (\ref{BCPF}) as 
\begin{eqnarray}
 \left.
  \left(
   \frac{\rm d}{{\rm d} y} {\rm Re} \Psi
   +
   c \sigma' {\rm Re} \Psi
  \right)
 \right|_{y = 0^+, \pi R^-}
 &\sim &  \left. -z \gamma^\mu\partial_\mu {\rm Re} \Psi_R \right|_{y = 0^+, \pi R^-}
 \nonumber\\
 &=& 
 \mp\frac{z \gamma^\mu\partial_\mu}2\sqrt{m_L^{0, \pi R}}^*\,{\psi_R^u}^{0, \pi R},
 \\
 \left. {\rm Im} \Psi \right|_{y = 0^+, \pi R^-}
 &=&
 \mp\frac12\sqrt{m_R^{0, \pi R}}^*\,{\psi_L^d}^{0, \pi R},
 \label{BCNonOrbifold}
\end{eqnarray}
 where the upper (lower) signs correspond to the boundary conditions at $y=0$ ($y=\pi R$). 
Because there are discontinuities of the wave function profiles at the boundaries 
 due to the mixing masses, we use limits, which are indicated by the superscripts $\pm$, 
 instead of the values on the boundary for the bulk fermions. 
Namely, for example, $0^+$ denotes $0+\epsilon$ with real and positive parameter 
 $\epsilon\to0$.
In the first line, we use the equation of motion to rewrite
 the left hand side by 
 the wave function of the right-handed partner, $\Psi_R$.
Strictly speaking, the derivative should 
 be the covariant one for this purpose. 
Nevertheless we can do it by the {\it non}-covariant derivative 
 because the difference does not contribute 
 to the KK mass function calculated below 
 (which is already mentioned above). 
We can remove the localized fermions in the boundary conditions using the equation 
 of motion for them, 
\begin{eqnarray}
 z \gamma^\mu\partial_\mu \psi_{L}^{0, \pi R} 
 &=& -\sqrt{m_R^{0, \pi R}} \left.{\rm Im}\Psi_R\right|_{y = 0, \pi R}, \\
 z \gamma^\mu\partial_\mu \psi_{R}^{0, \pi R} 
 &=& -\sqrt{m_L^{0, \pi R}} \left.{\rm Re}\Psi\right|_{y = 0, \pi R}. 
\end{eqnarray}
Then, we get 
\begin{eqnarray}
 z m_n \left.{\rm Im}\Psi\right|_{y = 0^+, \pi R^-} 
 &=&\pm\frac12 \left|m_R^{0, \pi R}\right| \left.{\rm Im}\Psi_R\right|_{y = 0, \pi R}, \\
 z m_n \left.{\rm Re}\Psi_R\right|_{y = 0^+, \pi R^-} 
 &=&\mp\frac12 \left|m_L^{0, \pi R}\right| \left.{\rm Re}\Psi\right|_{y = 0, \pi R}.
\end{eqnarray}
These give
 four conditions for four coefficients as before, 
 and we can calculate a KK mass function as 
\begin{eqnarray}
 N(x)
 &=&
 \left[\,
  \left(
   a x J_{c-1/2}(a x) - \left|\hat m_L^{0}\right| J_{c+1/2}(a x)
  \right)
  \left(
     x J_{-c+1/2}(x) - \left|\hat m_L^{\pi R}\right| J_{-c-1/2}(x)
  \right)
 \right.\nonumber\\
 &&\left.
 -\left(
     x J_{c-1/2}(x) + \left|\hat m_L^{\pi R}\right| J_{c+1/2}(x)
  \right)
  \left(
   a x J_{-c+1/2}(a x) + \left|\hat m_L^{0}\right| J_{-c-1/2}(a x)
  \right)
 \,\right] \nonumber\\
 &\times&
 \left[\,
  \left(
   a x J_{c+1/2}(a x) + \left|\hat m_R^{0}\right| J_{c-1/2}(a x)
  \right)
  \left(
     x J_{-c-1/2}(x) + \left|\hat m_R^{\pi R}\right| J_{-c+1/2}(x)
  \right)
 \right.\nonumber\\
 &&\left.
 -\left(
     x J_{c+1/2}(x) - \left|\hat m_R^{\pi R}\right| J_{c-1/2}(x)
  \right)
  \left(
   a x J_{-c-1/2}(a x) - \left|\hat m_R^{0}\right| J_{-c+1/2}(a x)
  \right)
 \,\right] \nonumber\\
 &-&
 \frac{2\cos^2(c\pi)}{a\pi^2 x^2}
 \left(
  x^2 + \left|\hat m_L^{0}\right| \left|\hat m_R^{0}\right|
 \right)
 \left(
  a^2 x^2 + \left|\hat m_L^{\pi R}\right| \left|\hat m_R^{\pi R}\right|
 \right)
 \left(
  \cos(2Q\theta_W)-1
 \right).  
\label{NonOrbifoldN}
\end{eqnarray}
This function coincides with Eq.~(\ref{Nc}) in the limit of 
 vanishing all mixing masses 
 up to an overall numerical factor and a factor
 $x^4$, which indicates 
 that there are four additional massless modes. 
These massless modes correspond to the four boundary fermions, 
 which we should remove in the case of vanishing mixing masses. 
If we introduce one of 
 $(m_L^0,m_R^0)$ while the other is zero, 
 the $\theta_W$-dependent
 term is suppressed by $1/m_{L,R}^0$ compared to other terms  
 when the mixing masses are huge enough. 
Thus, 
 the dependence of the KK mass function on $\theta_W$
 vanishes in the limit 
 $m_{L,R}^0\to\infty$. 
The same discussion can be 
 applied to a pair of $(m_L^{\pi R},m_R^{\pi R})$. 
Reminding that $m_L$ ($m_R$) changes the
 boundary condition of $\Psi^u$ ($\Psi^d$), 
 we can see that the dependence remains 
 if and only if the boundary conditions of $\Psi^u$ and $\Psi^d$ are 
 changed at the same time. 
Interestingly, in such cases, the boundary conditions return to
 orbifold-like ones.
In other words,
 for fermions with the non orbifold-like boundary conditions, 
 the KK mass function becomes independent of $\theta_W$,
 and thus such fermions 
 do not contribute to the effective potential of $\theta_W$. 

However, this conclusion seems strange, in a sense. 
It is because the coupling of the 
 non orbifold-like fermions to the Higgs field seems non-vanishing. 
For instance,  let us consider the flat limit $k\to0$. 
In this case, $\Psi^u$ component of the fermion with the first boundary condition 
 in Eq.~(\ref{NonOrbifoldBC}) interacts to the Higgs field with $\Psi^d_R$, which has 
 $(D,D)$ boundary condition. 
It is easy to see that the overlap integral among the zero mode of $\Psi^u$, 
 the lowest mode of $\Psi^d_R$, $\sin(y/R)$, and the zero mode of $A_5$ is 
 non-vanishing.
The reason 
 why the contributions from the non orbifold-like fermions vanish
 might be related to the fact
 that 
 a $(+,+)$ mode 
 ($\Psi^u$) does not couple to a $(-,-)$ mode ($\Psi^d_R$) through
 $\theta_W$
 in the orbifold picture.  
However, we do not have clear understanding at this stage, and 
 we leave this as an open question.

Before closing this section, let us comment on Eq.~(\ref{NonOrbifoldN}). 
In the above discussion, we use only the limits of  $m\to0$ or
$m\to\infty$ for
 examining the Neumann and Dirichlet boundary conditions. 
But, we can take finite value for the mixing mass, and in fact this is 
 often used in model building in the GH unification scenario 
 to make unwanted zero modes massive. 
The expression (\ref{NonOrbifoldN}) can be used for such cases.

\section{Gauge-Higgs Condition}
\label{Sec:GHC}

In this section, we examine the GH condition, proposed 
 in Ref.~\cite{GHC} in context of the GH unification scenario 
 in the flat background, in the warped GH unification.

The basic idea of the GH condition comes from the fact that, below the compactification scale, the effective theory should be described only by zero modes in usual four dimensional field theory. Since the Higgs field is merely a scalar field in the effective theory, its potential receives divergent corrections and need to be renormalized. The cutoff scale of the effective theory is expected to be around the compactification scale, and the theory is defined by renormalization conditions at that scale. We call the condition on the Higgs quartic coupling as the GH condition. Once we settle these conditions, we can make analysis in terms of a familiar four dimensional framework such as renormalization group equations, which is a powerful tool to investigate the low energy phenomena of the GH unification scenario.

\subsection{GH condition in the flat background}
\label{Sec:GHC-flat}

Before going to the discussion of the GH condition in the warped background, we briefly review the condition in the flat case. In Ref.~\cite{GHC}, we have investigated the effective potential of the Higgs field in the flat five dimensional space-time. Contributions to the effective potential of the Higgs field from periodic and anti-periodic fermions are analytically written as (up to constant term)
\begin{eqnarray}
 &&
 \sum_{w = 1}^{\infty} \frac{\cos(w x)}{w^5}
 =
 \zeta_R (5)
 -
 \frac{x^2}{2!} \zeta_R (3)
 +
 \frac{x^4}{4!} \frac{1}{2}
 \left[ \frac{25}{6} - \ln\left( x^2 \right) \right]
 +
 {\cal O}(x^6), 
 \label{exp1}
 \\
 &&
 \sum_{w = 1}^{\infty} \frac{\cos[w (x - \pi)]}{w^5}
 =
 -
 \frac{15}{16} \zeta_R (5)
 +
 \frac{x^2}{2!} \frac{3}{4} \zeta_R (3)
 -
 \frac{x^4}{4!} \ln (2)
 +
 {\cal O}(x^6)
 \label{exp2}
\end{eqnarray}
with the overall coefficient $C/2 = 3/(4\pi^2(2\pi R)^4)$. Additional minus signs appear for the contribution from bosons \cite{GeneralFormula}. Here $\zeta_R (x)$ is Riemann's zeta function and $x = Q g_4 2 \pi R h = m_1/\Lambda_{\rm UV}$, where $Q$ is the charge, $\Lambda_{\rm UV} = 2\pi R$ is the cutoff scale of the four dimensional effective theory and $m_1$ is the mass of the zero mode acquired after the symmetry breaking.

In the quadratic term of $x^2$, contributions from anti-periodic modes are the same order as those from periodic ones, and have opposite signs. Since anti-periodic fermions have no zero modes (the mass of the lightest mode is of the order of the cutoff scale), this fact means that the mass parameter can be treated as a free parameter as far as we are interested in the low energy effective theory in the GH unification scenario. Unlike the quadratic term, contributions come mostly from periodic modes in the quartic term when $x < 1$.

On the other hand, the effective potential is also calculated in the framework of the effective theory. The contribution to the quartic term from the zero mode of a periodic fermion with the charge $Q$ turns out to be
\begin{eqnarray}
 \left. V_{\rm eff} \rule{0mm}{0.4cm} \right|_{h^4}
 =
 \frac{1}{4!}
 \left[
  \lambda (\mu)
  +
  \frac{b}{2} \left( Q g_4 \right)^4
  \left\{
   \ln \left( \frac{h^2}{\mu^2 }\right) - \frac{25}{6}
  \right\}
 \right] h^4,
 \label{eff2}
\end{eqnarray}
where $\lambda(\mu)$ is defined by
\begin{eqnarray}
 \lambda(\mu)
 =
 \left. \frac{ {\rm d}^4 V_{\rm eff} }{ {\rm d}h^4 } \right|_{h = \mu},
 \label{RunningQuartic}
\end{eqnarray}
which is the renormalized coupling defined at the scale $\mu$ \cite{Coleman}, and $b = -3/\pi^2$ is the coefficient of the beta function of $\lambda(\mu)$ concerning the Yukawa coupling of the zero mode with the Higgs field. By comparing this result with Eq.(\ref{exp1}) (the spin degree of freedom is 4), we find the renormalization condition,
\begin{eqnarray}
 \lambda\left(\frac{1}{Qg_4L}\right)
 =
 \lambda\left(\frac{1}{Qg_42\pi R}\right)
 =
 0.
 \label{G-H condition}
\end{eqnarray}
This is the GH condition in the five dimensional model in the flat background. Since usually $Qg_4 = {\cal O}(1)$, the running coupling constant vanishes around the cutoff scale $\Lambda_{\rm UV}$. Practically, we can also find the cutoff scale $\Lambda_{\rm UV}$ from the fourth derivative of the effective potential around the origin, which is obtained from Eq.(\ref{exp1}) as
\begin{eqnarray}
 \left.\frac{{\rm d}^4V_{\rm eff}}{{\rm d}h^4}\right|_{h\to0}
 =
 \frac{b}2 y^4 \ln\left( \frac{m_1^2}{\Lambda_{\rm UV}^2}\right), 
 \label{cutoff}
\end{eqnarray}
where $y = Q g_4$ is the Yukawa coupling of the zero mode. Interestingly, this is nothing but the renormalization effect of the coupling from the cutoff scale $\Lambda_{\rm UV}$ down to the zero mode mass $m_1$, neglecting the Yukawa coupling flow.

\subsection{GH condition in the warped background}
\label{Sec:GHC-warp}

The GH condition is consistent with the physical speculation that the Higgs self interaction should vanish above the compactification scale where the five dimensional gauge invariance will be recovered. Therefore, it is expected that a similar condition holds also in the warped case. In that case, however, it is not clear which scale is the compactification scale. In fact, the typical scale of the first KK mass $m_{KK} = \pi ka$ is far from the 'radius' $1/R$. Unfortunately, it is difficult to investigate the effective potential analytically in the warped case. Instead, we make numerical analyses to find the cutoff scale $\Lambda_{\rm UV}$. Here, we examine the effective potential induced by a periodic mode with a charge $Q = 1/2$,
\begin{equation}
 V_{\rm eff}(\theta_W;c,1/2)
 =
 \frac12\frac{(ka)^4}{(4\pi)^2}
 2\int_0^\infty {\rm d} x x^3
 \ln \left( 1 + \frac{\cos\theta_W}{\bar{N_c}(x; c)} \right),
 \label{PotP1/2}
\end{equation}
where we fix the warp factor as $a = a_0 = 10^{-15}$. The contribution from an anti-periodic mode is given simply by $V_{\rm eff}(\theta_W+\pi;c,1/2)$.

\begin{figure}
 \begin{center}
  \scalebox{0.75}{\includegraphics*{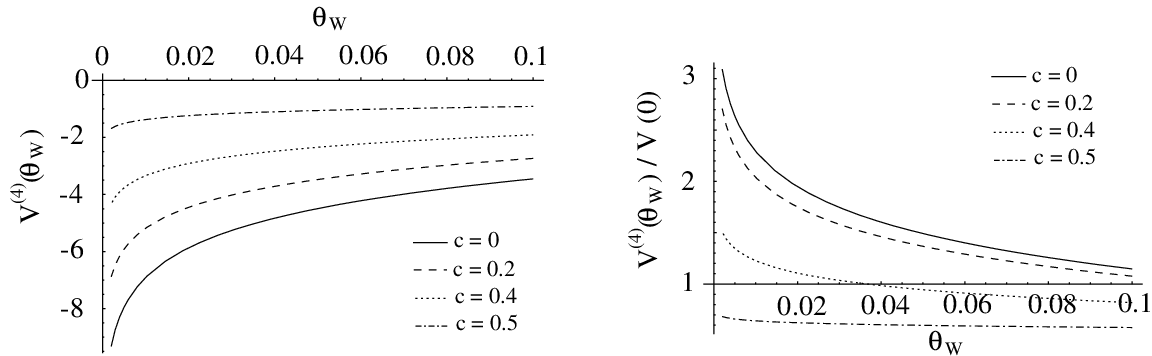}}
  \caption{\small The scale dependence of the running quartic coupling: The fermion mass is fix at $c=0,0.2,0.4,0.5$, respectively. In the left figure, the forth derivative of the potential is shown as a function of $\theta_W$, while it is normalized by $V_{\rm eff}(0;c,1/2)$ in the right figure.}
@\label{fig:GHC}
 \end{center}
\end{figure}

At first, let us see the scale dependence of the running quartic coupling defined in Eq.(\ref{RunningQuartic}). In Fig.\ref{fig:GHC}, we show the forth derivative of the potential, $V_{\rm eff}^{(4)}(\theta_W) \equiv{{\rm d}^4 V_{\rm eff} (\theta_W;c,1/2)} / {{\rm d} \theta_W^4}$, as a function of $\theta_W$. In the left figure, $V_{\rm eff}^{(4)}(\theta_W)$ is normalized by the prefactor $(ka)^4/(32\pi^2)$. Since
\begin{equation}
 \frac{{\rm d^4}}{{\rm d}h^4} 
 =
 \frac{g_4^4(1-a^2)^2\left(\ln(a)\right)^2}{4(ka)^4}
 \frac{{\rm d^4}}{{\rm d}\theta_W^4}, 
\end{equation}
the running coupling in Eq.(\ref{RunningQuartic}) is obtained by multiplying $2(\ln(a_0)/(16\pi))^2g_4^4 \sim 0.94g_4^4$. When $c$ close to 0, the coupling is so large that the perturbative calculation is not reliable. The reason why we show such unreliable results is to compare the result with reliable ones (with $c$ close to $1/2$). Since the potential in the $c = 0$ case is the same as that in the flat model, the GH condition is given by that in Sec.\ref{Sec:GHC-flat}. Therefore, with the correspondence, $1/R \leftrightarrow \pi ka/(1-a)$, the GH condition in Eq.(\ref{G-H condition}) holds at $\mu= ka/(2Qg_4(1-a))$ in the $c = 0$ case. The results in Fig.\ref{fig:GHC} show that the potential with smaller $c$ has the stronger IR divergence. In other words, the effect of the constant term in $V_{\rm eff}^{(4)}(\theta_W)$ becomes more relevant for the mode with $c$ closer to $1/2$. Thus, the scale of the GH condition is larger for larger $c$. It means that there is already non-vanishing threshold effect at $\mu= ka/(2Qg_4(1-a))$ which is the counterpart of the scale in the flat case. 

This fact seems to be natural, because, in the warped background, it is well known that the profiles of KK modes tend to approach to the IR ($y = \pi R$) brane, thus the modes have larger Yukawa couplings with the Higgs field (which also localizes around the IR brane) than the zero modes have. As a result, contributions from the KK modes are relatively enhanced, and their effects on the quartic coupling are expected to appear as threshold corrections. These threshold corrections from the KK modes at the compactification scale are sizable in the warped GH unification models, in contrast to the flat models.

\begin{figure}
 \begin{center}
  \scalebox{0.75}{\includegraphics*{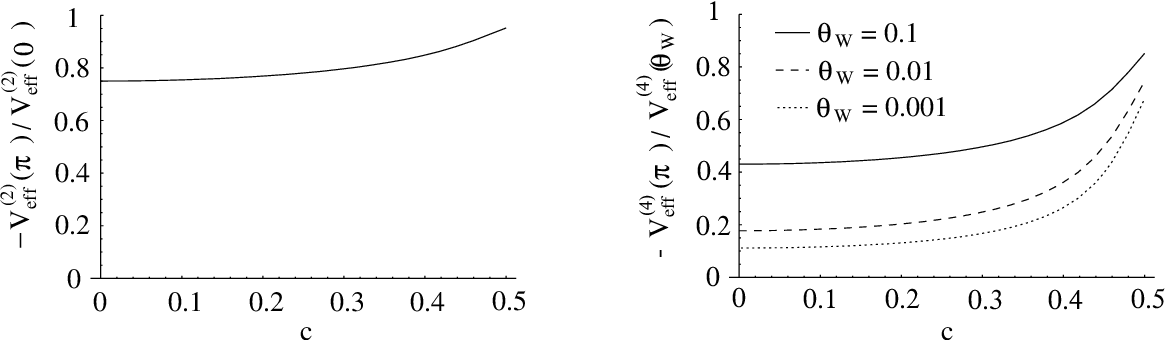}}
  \caption{\small Contributions to the potential from anti-periodic modes: In the left figure, the ratio between contributions from anti-periodic and periodic modes is shown. In the right figure, the Wilson phase is set to be $\theta_W=0.1, 0.01, 0.001$.}
  \label{fig:MassCorrection}
 \end{center}
\end{figure}

Next, we discuss contributions to the potential from the anti-periodic modes. The contributions to the quadratic term (mass parameter) are shown in Fig.\ref{fig:MassCorrection} (left figure) using the second derivative of the potential at $\theta_W = \pi$,
\begin{equation}
 V_{\rm eff}^{(2)}(\theta_W)
 \equiv
 \frac{{\rm d}^2 V_{\rm eff}(\theta_W;c,1/2)}{{\rm d} \theta_W^2}
 =
 - \frac12\frac{(ka)^4}{(4\pi)^2} 2\int_0^\infty {\rm d} x x^3
   \frac{1+\bar{N_c}(x;c) \cos\theta_W}
   {\left(\bar{N_c}(x;c)+\cos\theta_W \right)^2},
\end{equation}
normalized by $V_{\rm eff}^{(2)}(0)$ as a function of $c$. This ratio is nothing but that between anti-periodic and periodic modes. As can be seen in the figure, contributions from anti-periodic modes are comparable to those from periodic ones as in the flat case.

The contributions to the quartic term from anti-periodic modes are shown in Fig.\ref{fig:MassCorrection}, where the ratios $-V_{\rm eff}^{(4)}(\pi)/V_{\rm eff}^{(4)}(\theta_W)$  are depicted as a function of $c$ for $\theta_W = 0.1, 0.01, 0.001$. Notice that the potential with $c = 0$ is the same as that in the flat case. We see that contributions from anti-periodic modes are suppressed at small $\theta_W$ when $c \ll 1$ as expected. On the other hand, for $c$ close to $1/2$, the contributions are comparable to those from the periodic modes. One of the reasons should be the fact that the mass of the lightest mode of the anti-periodic fermion is much smaller than the typical KK mass $m_{KK}$. Also, the result is consistent with the fact that the threshold corrections from the KK modes are more important for larger $c$. The figure implies that the contribution from the running effect below the first KK mass is not large, which is different from the flat case. Unfortunately, it seems to be difficult to find an analytical expression for the threshold corrections from the KK modes.

\section{Summary and Discussions}
\label{summary}

In this article, we have derived formulae for calculating the effective potential of the Higgs field in the gauge-Higgs unification scenario on the Randall-Sundrum background. They can be applied even when we introduce bulk fermions with arbitrary parity-odd bulk mass terms and boundary conditions. These formulae will be useful not only for making analyses for the GH unification scenario but also for constructing realistic models having many attractive features \cite{AdS/CFT}-\cite{EWPT}. 

We have also calculated the contributions to the potential from bulk
 fermions with boundary conditions that are not allowed in the orbifold
 picture.
As a result, we have shown that their contributions vanishes 
 even though they seem to couple with the Higgs field. 
It might be related to the fact that the orbifold parity forbids 
 the coupling in the orbifold picture, but 
 we have not found a clear reason 
 why the contributions vanish and leave this as an open question. 
Anyhow we would notice that the expression of the KK mass function,  
 which is used to calculate the effective potential, 
 is useful for the finite values of the mixing 
 masses, although we derived it to examine the contribution of the non 
 orbifold-like fermions in the limit of zero or infinite mixing masses.

Since the formulae allow us to calculate the Higgs potential exactly at one-loop level, the potential properly incorporates its infrared behavior. Thus, we have examined the GH condition \cite{GHC} in the warped background. In the flat case, the running coupling of the quartic Higgs interaction vanishes at the compactification scale. On the other hand, in the warped models, we have found that the coupling have a substantial value already at the scale of the typical KK mass. This fact can be understood as the threshold corrections from the KK modes, because interactions of the Higgs with the KK modes are strong compared to those with zero modes. Though it is not easy to evaluate the threshold corrections analytically, once we find a way to evaluate those, it is possible to investigate GH models on the warped background in the framework of well established four dimensional field theory. We leave this problem as a future work. 

\section*{Acknowledgment}

T.~Y. would thank to K.~Oda for useful discussions. 
This work is supported in part by the Grant-in-Aid for Science Research, Ministry of Education, Science and Culture, Japan (No.16540258, No.17740146 for N.~H. and No.18740170 for N.~O.).
T.~Y. is supported in part by The 21st Century COE Program 
 ``Towards a New Basic Science; Depth and Synthesis''.


\end{document}